\documentclass[lettersize,journal]{IEEEtran}

\usepackage{amsmath,amsfonts}
\usepackage{array}
\usepackage[caption=false,font=normalsize,labelfont=sf,textfont=sf]{subfig}
\usepackage{textcomp}
\usepackage{stfloats}
\usepackage{url}
\usepackage{verbatim}
\usepackage{graphicx}
\usepackage{cite}

\usepackage[table]{xcolor}
\usepackage{booktabs}
\usepackage{multirow}
\usepackage{tabularx}
\usepackage{soul}
\renewcommand{\footnoterule}{%
  \kern -3pt
  \hrule width 2in height 0.4pt
  \kern 2.6pt
}
\usepackage[linesnumbered,ruled,vlined]{algorithm2e}
\SetNlSty{textbf}{}{:}
\SetKwInOut{KwIn}{Input}
\SetKwInOut{KwOut}{Output}

\begin{document}

\title{Modeling U.S. Attitudes Toward China via an Event-Steered Multi-Agent Simulator}

\author{Chenxu~Zhu, 
        Hantao~Yao,~\IEEEmembership{Member,~IEEE},
        Wu~Liu,~\IEEEmembership{Senior~Member,~IEEE}, 
        Junbo~Guo, 
        and~Yongdong~Zhang,~\IEEEmembership{Fellow,~IEEE}
\thanks{Chenxu Zhu, Hantao Yao, Wu Liu, and Yongdong Zhang are with the University of Science and Technology of China (USTC), Hefei 230026, China (e-mail: cx\_zhu@mail.ustc.edu.cn; yaohantao@ustc.edu.cn; liuwu@ustc.edu.cn; zhyd73@ustc.edu.cn).}
\thanks{Junbo Guo is with People.cn, Beijing, China (e-mail: guojunbo@people.cn).}
\thanks{Corresponding author: Yongdong Zhang.} 
}
\maketitle
\begin{abstract}
Understanding the dynamic evolution of opinions, such as U.S. public attitudes toward China, is essential for assessing geopolitical risks. 
However, existing LLM-based multi-agent simulators predominantly rely on static rules and fixed datasets, limiting their ability to capture the dynamic, event-driven nature of macro-level opinion shifts in real-world settings.
To address this limitation, we propose an Event-Steered Multi-Agent Simulator (ES-MAS), in which significant events and daily news continuously drive opinion evolution through dynamic interactions among agents.
We first construct the China–U.S. Relation Evolution (CURE) dataset, covering 20 quarters from 2021 to 2025, including 258 major events and over 14,000 daily news articles, and providing a comprehensive temporal foundation for modeling opinion dynamics.
Building upon the CURE dataset, we propose a Dual-Stream Data Integration Engine (DSDIE) that aligns simulations with historical timelines via macro-level events while enabling personalized information exposure based on individual agent profiles and contextual signals. 
Furthermore, we design a News-Driven Dynamic Interaction (NDDI) module, which adaptively groups agents with shared news interests into localized interaction contexts, facilitating bottom-up consensus formation while mitigating the risk of isolated information cocoons.
Experimental results on the CURE dataset demonstrate that ES-MAS substantially outperforms existing simulators in reproducing real-world historical trends, offering a scalable and effective framework for modeling dynamic opinion evolution.
\end{abstract}


\section{Introduction}
\noindent
The dynamic evolution of opinions in social networks plays a crucial role in shaping collective behavior and informing policy decisions. 
Understanding these processes is essential for anticipating societal trends and mitigating potential risks. 
In this context, China--U.S. relations represent a major geopolitical challenge of the 21st century, with significant implications for global stability~\cite{allison2017destined}. 
Public attitudes toward this relationship are highly dynamic, influenced by events, media narratives, and social interactions. 
Ignoring these shifts may lead to misinformed decisions and increased risks of conflict~\cite{li2025media}. 
Therefore, accurately analyzing and forecasting U.S. public attitudes toward China is critical. 
However, existing approaches mainly rely on surveys and expert analyses~\cite{campbell2018china}, which are resource-intensive and insufficient for capturing the rapid, continuous evolution of real-world public sentiment. 
\begin{figure}
    \centering
    \includegraphics[width=1\linewidth, height=5.8cm]{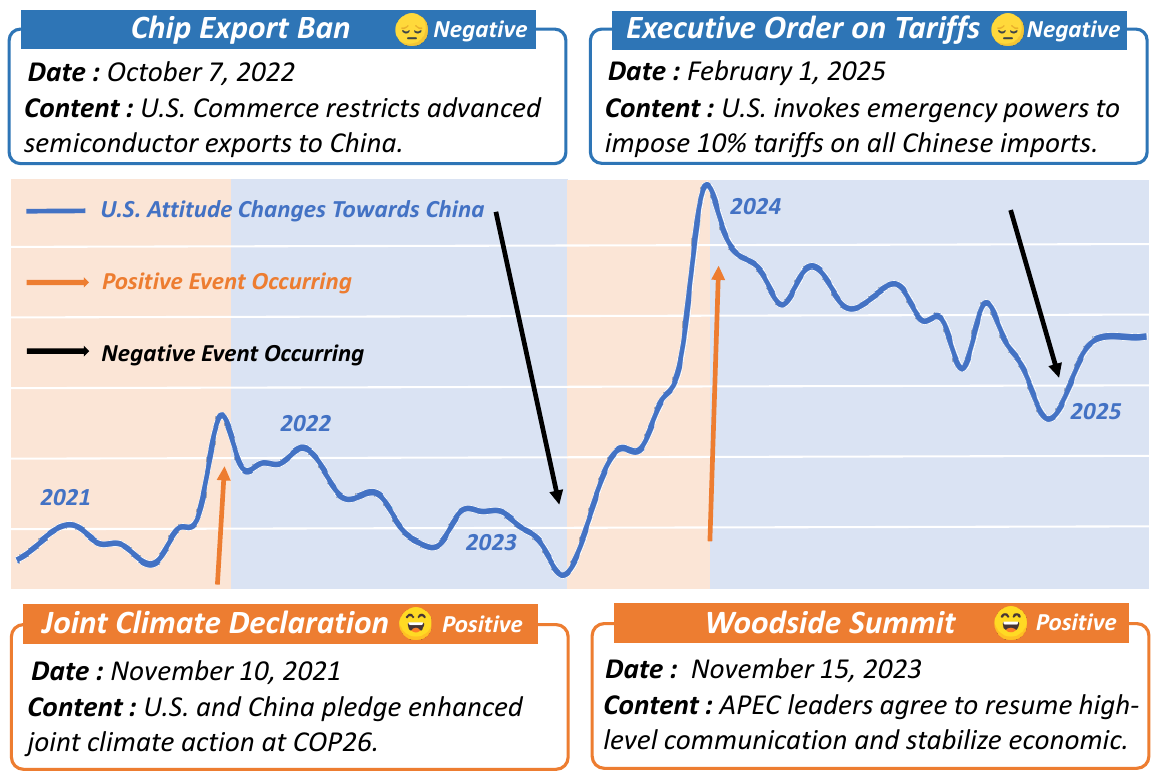} 
    \caption{The trajectory of U.S. attitudes toward China, illustrating how Real-world Geopolitical Events shape macro-level trends }
    \label{fig:attitude} 
\end{figure}

Unlike traditional approaches that rely on human expertise for analysis~\cite{silver2023americans}, there has been growing interest in leveraging machine learning and artificial intelligence to study the evolution of U.S. attitudes toward China~\cite{wang2023artificial}. 
In particular, Agent-Based Models (ABMs) have been widely adopted to simulate macro-level social dynamics through the collective interactions of individuals~\cite{deffuant2000mixing,deffuant2002can}. 
However, these models typically rely on simplified heuristic rules, failing to capture the heterogeneity and cognitive complexity of human decision-making~\cite{lazer2009computational}. 
Recent advances in large language models (LLMs) have enabled the development of agent-based simulators with enhanced reasoning and decision-making capabilities, allowing agents to interpret complex geopolitical contexts~\cite{hua2023war,guan2024richelieu}, providing new opportunities for modeling individual-level perceptions of China-U.S. relations. 
Nevertheless, existing LLM-based simulators often adopt static interaction protocols and depend on pre-collected datasets to model public sentiment and its evolution~\cite{mou2024unveiling,gao2023s3}. 
Such designs lack continuous external inputs, limiting their ability to capture the dynamic and event-driven nature of bilateral relations. 
Moreover, some recent work incorporates streaming daily news to infer societal trends by aggregating independently generated opinions from large populations of agents~\cite{sukiennikdebiasing}. 
However, by overlooking peer-to-peer interactions, these approaches fail to model the social diffusion of opinions and tend to produce ``information cocoons'', where agents remain confined to isolated perspectives. 
In conclusion, existing simulators remain inadequate for accurately reproducing the systemic and dynamic evolution of China-U.S. relations. 
Therefore, it is essential to consider both real-time data stream integration and the evolution of dynamic interactions to accurately simulate U.S. attitudes toward China.

A compelling motivation is to simulate macro-level bilateral trends through the lens of micro-level individual opinions, employing the integration of meaningful real-world data between China and the United States alongside dynamic multi-agent interactions~\cite{schelling2006micromotives,degroot1974reaching}. 
The simulation framework is grounded in dual information streams that incorporate both significant macro-level events and micro-level daily news. 
Unlike traditional simulation algorithms that depend exclusively on daily news data, we contend that substantial events influencing societal dynamics between China and the United States are essential for accurately modeling U.S. attitudes toward China. 
For example, the escalation of the Chip Export Ban in 2022 led to a significant deterioration in bilateral relations, while the leadership meeting at the Woodside Summit in 2023 contributed to a reduction in tensions, as illustrated in Figure~\ref{fig:attitude}. 
Within the simulation process, dynamic interactions among agents—driven by real-time news—effectively translate micro-level individual opinions into macro-level U.S. attitudes toward China. 
Research on small-world networks suggests that individuals are more likely to interact with peers who share similar opinions~\cite{mcpherson2001birds}. 
Furthermore, the dissemination of opinions can be achieved through the interconnectedness of numerous small-world networks on a global scale~\cite{watts1998collective}. 
Through the analysis of news content, individuals with analogous views on China-U.S. relations can be clustered into distinct groups, allowing for dynamic interactions that facilitate the evolution of cognitive perspectives. 
Simultaneously, interactions between these groups can foster the spread of opinions on a broader level, enabling the deduction of the overall macro situation.

In light of the aforementioned motivation, we introduce a novel Event-Steered Multi-Agent Simulator (ES-MAS) designed to model U.S. attitudes toward China by utilizing significant real-time events and dynamic interactions between agents.
To ensure that our simulations reflect real-world dynamics, we first develop the  \textbf{C}hina-\textbf{U}.S. \textbf{R}elation \textbf{E}volution (CURE) dataset, which comprises significant events and daily news related to China-U.S. interactions over the past five years. 
Next, we introduce a Dual-Stream Data Integration Engine to incorporate both significant events and daily news into the simulator. 
At the macro level, significant events establish the overarching historical context of China-U.S. relations. 
At the micro level, a Personalized Autonomous Ingestion Mechanism (PAIM) is proposed to select daily news based on agent preferences and current events, promoting cognitive diversity.
Additionally, we propose a News-Driven Dynamic Interaction (NDDI) module that uses news topics as signals to dynamically cluster agents who are interested in the same China-U.S. news. 
This clustering process creates locally relevant interaction contexts, allowing agents to engage in group-level discussions.
Overall, this strategy effectively demonstrates how micro-level interactions can influence the macro-evolution of China-U.S. relations from the bottom up.

Experiments conducted on the \textbf{CURE} dataset show that {ES-MAS} achieves superior performance in simulating U.S. Attitude Changes Toward China compared to existing social simulators. 
In summary, the major contributions can be summarized as follows:
\begin{itemize}
    \item We propose a Dual-Stream Data Integration Engine (DSDIE) to leverage significant events and daily news to effectively bridge macroscopic trends with micro-individual cognition.
    \item We propose a News-driven Dynamic Interaction (NDDI) module to dynamically cluster the agents into several groups for conducting the intra-group interaction.
    \item We construct a novel \textbf{C}hina-\textbf{U}.S. \textbf{R}elation \textbf{E}volution (CURE) dataset comprising significant events and daily news related to China-U.S. interactions over the past five years.
\end{itemize}

\section{Related Work}
\textbf{Opinion Dynamics Modeling:}
Opinion dynamics research links micro-level interactions to macro-level phenomena via frameworks like selective exposure~\cite{frey1986recent} and social identity theory~\cite{tajfel2004social}. 
Traditional surveys offer empirical grounding but remain labor-intensive and static~\cite{lazer2009computational}. 
To capture dynamic processes, modeling approaches have evolved from probabilistic methods~\cite{degroot1974reaching} to traditional Agent-Based Models (ABMs); however, these ABMs typically simulate social dynamics using heuristic rules and often fail to capture human cognitive complexity~\cite{lorenz2021individual,berry2002adaptive}. 
In contrast, current studies demonstrate that Large Language Models (LLMs) can reproduce complex cognitive behaviors, including human-like polarization on political issues~\cite{tornberg2023simulating}, in-group favoritism~\cite{argyle2023out}, and cognitive dissonance~\cite{aher2023using}.  
Furthermore, the role of external information environments, such as mass media and critical exogenous events, is pivotal in shaping these cognitive trajectories~\cite{zaller1992nature}, our work employs dynamic social interactions within a population of agents to simulate macroscopic trends in international perceptions.

\textbf{LLM-based Social Simulators:}
LLMs empower agents to exhibit human-like behaviors, self-awareness ~\cite{xi2025rise,wang2024survey} and dispute deliberation and collective decision-making~\cite{sun2026cyberjurors}. 
For example, Generative Agents~\cite{park2023generative} simulate the daily interactions among 25 agents in a virtual town, demonstrating that LLM-based agents can generate social behaviors indistinguishable from human-created content. 
Recent simulator have drastically scaled these societies; for instance, frameworks like OASIS~\cite{yang2024oasis} and AgentSociety~\cite{piao2025agentsociety} achieve massive-scale simulations to study emergent social phenomena such as emotional contagion, and misinformation spread.
Building on this foundation, some multi-agent simulators utilize static data and fixed topologies for sociological analysis~\cite{liu2024skepticism,mou2024unveiling}, while recent graph-based frameworks such as ~\cite{zhou2026gasim} and ~\cite{sun2025dynamix} further explore large-scale dynamic social networks and evolving interaction structures. 
In the domain of opinion dynamics, other works incorporate time-varying data streams to track opinion changes, focusing on updating individual agent beliefs based on external information~\cite{chuang2024simulating,park2022social}. 
Another category of research targets macro-level political outcomes, such as elections or the evolution of international relations~\cite{DBLP:journals/corr/abs-2410-20746}, by modeling collective opinion dynamics to forecast systemic developments. 
In this work, ES-MAS introduces a dual-stream data integration engine and dynamic interactions among agents to synchronize the simulated society with historical reality.

\section{China-U.S. Relation Evolution Dataset}
To accurately simulate U.S. attitudes change toward China, we collect a novel \textbf{C}hina-\textbf{U}.S. \textbf{R}elation \textbf{E}volution (CURE) dataset consisting of the significant events and daily news related to China-U.S. interactions over the past five years. 

\begin{figure*}
    \centering
    \includegraphics[width=1\linewidth]{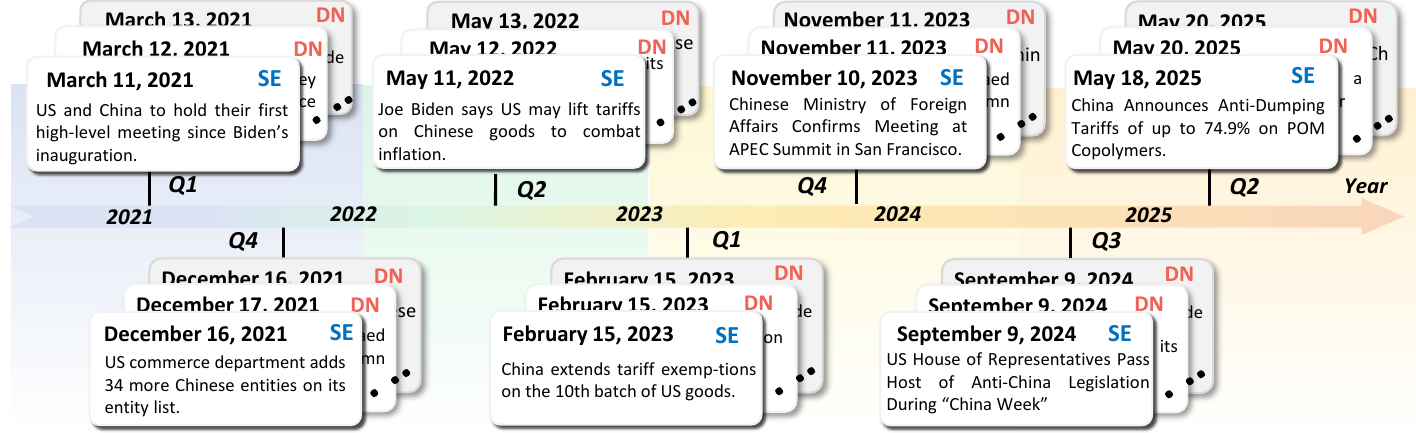}
    \caption{The content and structure of the CURE dataset about the Significant Events (SE) and Daily News (DN)}
    \label{fig:data}
\end{figure*}

\textbf{Significant Events (2021--2025):} 
Significant events play a crucial role in shaping U.S. attitudes toward China. 
To better understand this, we collected significant events from both the Biden and Trump administrations, sourced from \emph{China Briefing}\footnote{\url{https://www.china-briefing.com/}}, a policy intelligence platform. 
This documentation includes important occurrences such as presidential summits, trade sanctions, technology sector restrictions, and diplomatic conflicts.
In total, we collected 258 significant events from 2021 to 2025. 
For example, one event is: ``\emph{May 12, 2025: China and the U.S. agree to reduce reciprocal tariffs to 10\%.}" 
This event illustrates the complexities of the relationship between China and the United States.

\textbf{Daily News (2021--2025):} 
Besides the significant events, we also collect many daily news items related to China-U.S..
Specifically, we utilized \textit{public media API interfaces}\footnote{\url{https://open-platform.theguardian.com/}} to collect daily news spanning the periods of the Biden administration and the re-emergence of the Trump administration. 
To ensure comprehensiveness, we retrieved articles containing the keywords ``China" and ``U.S." across distinct categories, including politics, economy, technology, and environment. 
Finally, we collected over 30,000 articles of daily news from 2021 to 2025.
However, these daily news items are not directly relevant to shifts in the U.S. attitude towards China. 
To remove the noise from all articles, we employed the Qwen3-8B to justify whether the article contributes to the evolution of China-U.S. relations. 
For example, one news reads: ``\emph{May 12, 2025: U.S. Senate passes Taiwan UN resolution, sparks China debate.}".
Following the pre-processing, over 14,000 relevant articles remained.

    


By combining the collected significant events and daily news, we obtain the \textbf{C}hina-\textbf{U}.S. \textbf{R}elation \textbf{E}volution (CURE) dataset.
Each term of the CURE dataset consists of the \emph{Data}, \emph{Significant Event}, and \emph{Daily News}.
The detailed content and integration structure of the CURE dataset are illustrated in Figure~\ref{fig:data}.

\textbf{Dataset Statistics:}
Spanning 20 quarters from 2021 to 2025, the CURE dataset contains a total of 258 Significant Events (SE) and 14,650 Daily News (DN) articles. 
As detailed in Table~\ref{tab:detailed_dist}, the data distribution reflects the real-world intensity of U.S.-China relations. 
For example, the frequency of significant events peaks sharply in 2025 (90 events), aligning with the policy shifts and geopolitical friction during the U.S. administration transition. 
Meanwhile, the daily news, averaging over 2,900 articles annually, provides a dense background narrative.

\begin{table}
    \centering
    \caption{Detailed Quarterly Distribution of Significant Events (SE) and Daily News (DN).}
    \label{tab:detailed_dist}
    
    \renewcommand{\arraystretch}{1.2} 
    
    \setlength{\tabcolsep}{3pt}        
    
    \resizebox{0.98\linewidth}{!}{%
        \begin{tabular}{c|cc|cc|cc|cc|cc}
        \hline
        {Year} & \multicolumn{2}{c|}{{Q1}} & \multicolumn{2}{c|}{{Q2}} & \multicolumn{2}{c|}{{Q3}} & \multicolumn{2}{c|}{{Q4}} & \multicolumn{2}{c}{{Total}} \\
        & {SE} & {DN} & {SE} & {DN} & {SE} & {DN} & {SE} & {DN} & {SE} & {DN} \\
        \hline
        \hline
        2021 & 8  & 527 & 15 & 1,040 & 7  & 482 & 17 & 1,190 & 47 & 3,239 \\
        2022 & 7  & 732 & 9  & 937  & 7  & 707 & 5  & 516 & 28 & 2,892 \\
        2023 & 7  & 328 & 19 & 891  & 14 & 640 & 19 & 928 & 59 & 2,787 \\
        2024 & 10 & 742 & 7  & 470  & 9  & 696 & 8  & 572 & 34 & 2,480 \\
        2025 & 19 & 675 & 32 & 1,100 & 22 & 798 & 17 & 679 & 90 & {3,252} \\
        \hline
        \textbf{Sum} & \textbf{51} & \textbf{3,004} & \textbf{82} & \textbf{4,438} & \textbf{59} & \textbf{3,323} & \textbf{66} & \textbf{3,885} & \textbf{258} & \textbf{14,650} \\
        \hline
        \end{tabular}%
    }
\end{table}
\textbf{Ground Truth Attitudes:}
To effectively evaluate whether the evolution of U.S. attitude towards China trajectories generated by our simulator align with historical facts, we give a ground truth of U.S. attitude towards China based on the \textit{Database of Relations Between China and Major Powers} (1950--2025)\footnote{\url{https://www.tuiir.tsinghua.edu.cn/}} released by the Institute of International Relations at Tsinghua University (TUIIR). 
Note that the TUIIR database offers monthly updates, enabling it to capture subtle shifts in China-U.S. relations, such as the transitions between confrontation and detente during the Biden and Trump administrations. 
Finally, we extracted TUIIR scores from 2021 to 2025, ensuring temporal alignment for our simulation with reality.
\begin{figure*}
    \centering
    \includegraphics[width=0.95\linewidth]{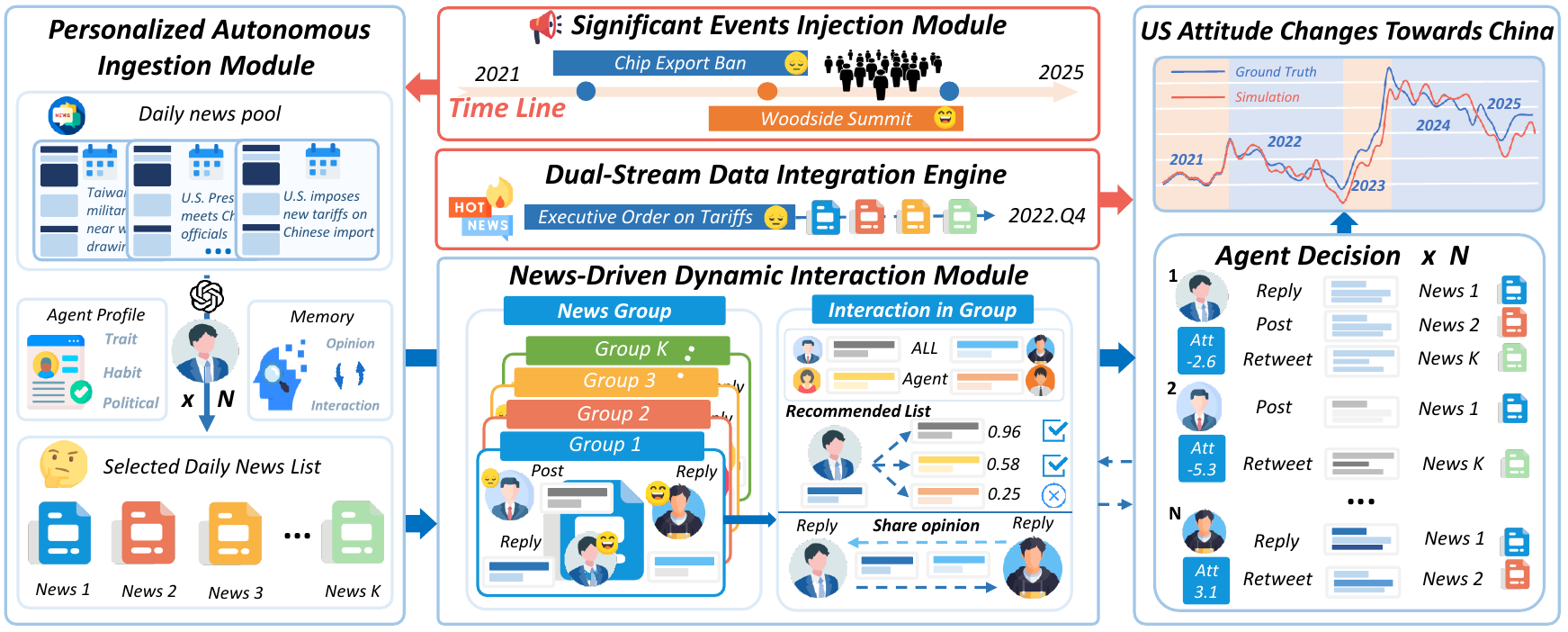} 
    \caption{The architecture of {ES-MAS} framework. Simulating U.S. Attitude Changes Toward China by Dual-Stream Data Integration Engine and News-Driven Interaction Module. 
    The engine integrates significant events and daily news to generate individual attitudes and daily news-responsive actions. 
    Subsequently, these actions facilitate the formation of news-interest groups, where dynamic interactions occur to influence the attitude evolution for the next step.}
    \label{fig:framework} 
    
\end{figure*}

\section{Event-Steered Multi-Agent Simulator}

Based on the CURE dataset, we introduce a novel Event-Steered Multi-Agent Simulator (ES-MAS), in which the opinions of multiple agents evolve under the influence of significant events and daily news.  
As shown in Figure~\ref{fig:framework}, ES-MAS comprises two components: a Dual-Stream Data Integration Engine and a News-Driven Dynamic Interaction Module.  
The Dual-Stream Data Integration Engine synchronizes simulations with historical timelines through macro-level significant events while also allowing individualized information intake at the agent level.  
The News-Driven Dynamic Interaction (NDDI) Module dynamically groups agents according to news content and models the evolution of collective attitudes. 
Its strength lies in effectively demonstrating how micro-level local interactions can drive the macro-level evolution of bilateral relations from the bottom up.  

Formally, ES-MAS is modeled as a multi-agent system with $N$ agents, denoted as $ \mathbb{S} = \{ \mathbb{A}_1, \mathbb{A}_2, \dots, \mathbb{A}_N \} $.  
Each agent configuration $\mathbb{A}_i$ consists of the agent $a_i$, the persona profile $\mathcal{P}_i$, and the memory $\mathcal{M}_i$; that is, $\mathbb{A}_i = \{ a_i, \mathcal{P}_i, \mathcal{M}_i \}$.  
Note that each agent $a_i$ is initialized with a multi-dimensional persona profile $ \mathcal{P}_i$, constructed based on demographic distributions~\cite{DBLP:journals/corr/abs-2410-20746,sukiennik2025roots}, and includes attributes such as name, age, gender, interests, personality, and political views.  
In addition to the profile, the agent’s memory $\mathcal{M}_{i,t}$ initialized at simulation step \( t = 0 \) serves as another key component for adapting the agent’s opinions.  

In our simulation from 2021 to 2025, ES-MAS operates on quarterly (three-month) time steps, corresponding to discrete time steps $t \in \{1, 2, \dots, 20\}$.  
At each step t, each agent $a_i$ receives dual-stream news $ \mathcal{D}_{i,t}$ generated by the Dual-Stream Data Integration Engine(DSDIE).  
Based on its persona $\mathcal{P}_i$ and the received news $\mathcal{D}_{i,t}$, each agent produces two outputs:  
1) An attitude score $ O_{i,t} \in [-10, 10] $, reflecting sentiment toward China;  
2) A news-responsive action $\mathcal{B}_{i,t}$, \emph{e.g.,} posting, retweeting, or replying, representing behavioral engagement with the news stream.  
Through the News-Driven Dynamic Interaction (NDDI) module, these responses $\mathcal{B}_{i,t}$ facilitate dynamic inter-agent interactions and promote group formation based on shared opinions. These interactions constitute the micro-foundations underlying the evolution of U.S.–China relations.  
Finally, by aggregating the individual attitude scores $O_{i,t}$ across all agents, we derive the macro-level trajectory of U.S. attitude changes toward China.
\enlargethispage{\baselineskip}
\vspace{-4mm}
\subsection{Dual-Stream Data Integration Engine}
To ensure the simulation aligns with real-world dynamics, we propose the Dual-Stream Data Integration engine to integrate information streams by jointly considering the significant events and daily news.
Formally, we construct a comprehensive dual-stream data $\mathcal{D}_{i,t}$,
\begin{equation}
\mathcal{D}_{i,t} = \{\mathcal{S}_{t} , \mathcal{I}_{i,t} \},
\end{equation}
where $\mathcal{S}_{t}$ is the significant events steam, and $\mathcal{I}_{i,t}$ is the personalized autonomous daily news stream.
For the agent $a_i$, the Significant Events Inject Module(SEIM) is proposed to generate the event data $\mathcal{S}_{t}$, and the Personalized Autonomous Ingestion Module(PAIM) is proposed to select the personalized daily news $\mathcal{I}_{i,t}$. 
In the following, we give a detailed description of those two components.

At the macro-level, SEIM exposes all agents $\{{a}_i\}_{i=1}^{N}$ to the significant events before each simulation step, effectively synchronizing the simulated society with the historical reality of China-U.S. relations.
Formally, for each simulation step, the set of significant events is defined as $\mathcal{S}_{t} = \{s_{t,1}, s_{t,2}, \ldots, s_{t,L}\}$, where $L$ denotes the number of events in the current quarter, and each event contains the event ID, title, date, and content.
 

Different from SEIM, the Personalized Autonomous Ingestion module (PAIM) constructs a candidate news pool derived from the daily news dataset, enabling agents to autonomously select daily news in this pool that align with their interests. 
Specifically, based on the current significant events $\mathcal{S}_{t}$, the agent utilizes its persona profile $\mathcal{P}_{i}$ and recent memory $\mathcal{M}_{i,t}$ as decision criteria to ingest relevant articles $\mathcal{I}_{i,t}$ from the candidate news pool $\mathcal{W}_{t} = \{w_1, w_2, \dots, w_\mathcal{N}\}$ with the Large Language Model(LLM),
\begin{equation}
\mathcal{I}_{i,t} = LLM \left( \mathcal{F}_{ingest}( \mathcal{S}_{t}, \mathcal{W}_{t}, \mathcal{P}_i, \mathcal{M}_{i, t}) \right),
\end{equation}
where $\mathcal{F}_{ingest}(\cdot)$ functions as the selection engine retrieve the top-$K$ news items that exhibit the semantic alignment with the agent's persona, memory, and environment. More details are provided in the Appendix.
\enlargethispage{\baselineskip}

Finally, the dual-stream data $\mathcal{D}_{i,t}$ establishes a stable macro-level cognitive foundation for the simulation, facilitating diverse micro-level cognitive processing. This dual-stream architect guarantees that the simulation of U.S. attitude changes toward China evolves dynamically within the boundaries of historical facts.
\subsection{News-Driven Dynamic Interaction Module}
In multi-agent simulations, the dynamic interactions between agents play a crucial role in deriving societal behavior at the macro level from individual actions at the micro level~\cite{tang2025gensim,piao2025agentsociety}. 
For example, we can infer the global attitude of China and the United States based on individual attitudes toward both countries. 
Inspired by small-world networks, people communicate primarily with a few individuals they are connected to within the small social network. 
Therefore, in multi-agent simulations, agents often interact dynamically with several agents belonging to the same group. 
Moreover, multiple groups can achieve global-level dynamic interactions through shared agents. 
Unlike existing simulators using a fixed topology or social topology, we construct a local interaction topology by considering agents who focuses the same daily news. 
To facilitate this, we propose the News-Driven Dynamic Interaction (NDDI) module to simulate interactions among agents based on dual-stream events, capturing the complex interplay between overarching bilateral trends and public interactions.

Given the dual-stream dataset $\mathcal{D}_{i,t}=\{\mathcal{S}_{t} , \mathcal{I}_{i,t} \}$ at the $t$-th step, it contains $K$ circulating news items, \emph{i.e.,} $\mathcal{I}_{i,t} =\{\mathcal{I}_{i,t,1},\mathcal{I}_{i,t,2},.......,\mathcal{I}_{i,t,K}\}$, where $\mathcal{I}_{i,t,k}$ is the $k$-th terms.
At each step, daily news $\mathcal{I}_{i,t}$ and significant event $\mathcal{S}_{t}$ give the agent for selectively engaging with this news by performing actions (\textsc{Post}, \textsc{Retweet}, or \textsc{Reply}). 
Inspired by real social interaction, all agents acting on the same news item are clustered into a corresponding news-driven group. 
For the given news item $\mathcal{I}_{i,t,k}$, the corresponding group $\mathcal{G}_{t,k}$ comprises the agents who are interested in the same daily news,
\begin{equation}
\mathcal{G}_{t,k} = \{ a_i \in \mathcal{A} \mid \mathcal{E}(a_i, \mathcal{I}_{i,t,k})=1 \},
\end{equation}
where $\mathcal{E}(a_i, \mathcal{I}_{i,t,k})=1$ denotes that the agent $a_i$ is intereseting on the daily news $\mathcal{I}_{i,t,k}$, and $\mathcal{A}$ is the set of all agents.
Finally, we group all agents $\mathcal{A}=\{a_1, a_2,...,a_{N}\}$ into $K$ groups $\mathbb{G}_{t}=\{\mathcal{G}_{t,1},\mathcal{G}_{t,2},...,\allowbreak\mathcal{G}_{t,K}\}$. More details are provided in the Appendix.
\begin{algorithm}
    \caption{Event-Steered Multi-Agent Simulator (ES-MAS)}
    \label{alg:es_mas}
    \KwIn{
        Agent Set $\mathbb{S}$; Steps $T$; Significant Events $\mathcal{S}_{t}$; Candidate News Pool $\mathcal{W}_{t}$.
    }
    \KwOut{
        Agents' attitude $\{O_{i,t}\}_{t=1}^T$ and behaviors $\{B_{i,t}\}_{t=1}^T$ for all agents $i \in \{1, \dots, N\}$.
    }
    \Begin{
        Initialize agent configurations $\mathbb{S} = \{ \mathbb{A}_1, \mathbb{A}_2, \dots, \mathbb{A}_N \}$, $\mathbb{A}_i = \{ a_i, \mathcal{P}_i, \mathcal{M}_{i,0} \}$ and initial attitude ${O}_{i,0}$ for $i=1 \dots N$\;
        
        \For{each timestep $t = 1$ \KwTo $T$}{
            Expose $\mathcal{S}_t$ for each agent\;
    
            \For{each agent $\mathbb{A}_i \in \mathbb{S}$}{
                Select news $\mathcal{I}_{i,t}$ via PAIM based on $\mathcal{S}_t, \mathcal{W}_{t}, \mathcal{P}_i$ and $\mathcal{M}_{i,t-1}$\;
                Integrate data $\mathcal{D}_{i,t} = \{ \mathcal{S}_t, \mathcal{I}_{i,t} \}$\;
                Generate initial attitude $O_{i,t}$ and behavior $B_{i,t}$ via $\mathcal{F}_{act}$ based on $\mathcal{D}_{i,t}, \mathcal{P}_i, \mathcal{R}_{i,t,k}$ and $\mathcal{M}_{i,t-1}$\;
            }   
            Partition $\mathbb{S}$ into groups $\mathbb{G}_t = \{ \mathcal{G}_{t,1}, \dots, \mathcal{G}_{t,K} \}$ based on actions $B_{i,t}$\;
    
            \For{each group $\mathcal{G}_{t,k} \in \mathbb{G}_t$}{
                \For{each agent $a_i \in \mathcal{G}_{t,k}$}{
                    Form reference stream $\mathcal{R}_{i,t,k}$ using visible replies $\mathcal{C}_{i,t,k}$ and top-$\tau$ filtered peers $\mathcal{V}_{i,t,k}$\;
                }
            }
            Attitude and memory update of the Environment ;
        }
        \Return $\{O_{i,t}, B_{i,t}\}_{i=1 \dots N}^{t=1 \dots T}$\;
    }
    
\end{algorithm}

After that, given the daily news $\mathcal{I}_{i,t,k}$, the agent $a_i$ belonging to the group $\mathcal{G}_{t,k}$ conducts the interaction to produce the attitude towards China $O_{i,t}$, behavioral response $\mathcal{B}_{i,t}$, and updated memory $\boldsymbol{M}_{i,t}$,
\setlength{\abovedisplayskip}{8pt}
\setlength{\belowdisplayskip}{8pt}
\begin{equation}
\small
O_{i,t}, \mathcal{B}_{i,t}, \boldsymbol{M}_{i,t} = LLM  (\mathcal{F}_{act} \left(\mathcal{D}_{i,t}, \mathcal{P}_i, \boldsymbol{M}_{i, t-1}, \mathcal{R}_{i,t,k} \right)), 
\label{eq:4}
\end{equation} 
where $\mathcal{P}_i$ is the personal profile, and $\mathcal{F}_{act}(\cdot)$ is the decision engine to generate opinion and news-responsive actions, manifesting as operations including \textsc{Post}, \textsc{Retweet}, and \textsc{Reply}.
$\mathcal{R}_{i,t,k}$ is the reference information stream for an agent $a_i$ when interacting within a group $\mathcal{G}_{t,k}$. More details are provided in the Appendix.

When the agent conducts the interaction with Eq.~\eqref{eq:4}, a critical problem is how to collect the reference information stream $\mathcal{R}_{i,t,k}$ among the agents from the group $\mathcal{G}_{t,k}$. 
In the News-Driven Dynamic Interaction, the act of engaging with the same news item establishes a shared information set to make peer actions potentially visible.
Moreover, an agent would refer to other agents' responses to the same news to decide its own actions.
Therefore, the reference information stream $\mathcal{R}_{i,t,k}$ consists of two terms,
\setlength{\abovedisplayskip}{8pt}
\setlength{\belowdisplayskip}{8pt}
\begin{equation}
\mathcal{R}_{i,t,k} = \mathcal{C}_{i,t,k} \cup \mathcal{V}_{i,t,k},
\end{equation}
where $\mathcal{V}_{i,t,k}$ is the dynamic information stream of agent $a_i$ from other agents, and $\mathcal{C}_{i,t,k}$ denotes the set of observable reply contents to news item $\mathcal{I}_{i,t,k}$ at time $t$ that are visible to agent $a_i$,
\begin{equation}
    \mathcal{C}_{i,t,k} = \left\{ \mathcal{B}_{j,t}^{\textsc{Reply}} \;\middle|\; a_j \in \mathcal{G}_{t,k}, \; j \neq i \right\},
\end{equation}
where $\mathcal{B}_{j,t}^{\textsc{REPLY}}$ denotes the agent $a_j$ conducts the same ``REPLY'' operation as agent $a_i$.

    


As mentioned above, an agent would refer to other agents' responses to the same news to decide its own actions.
For example, when agent $a_i$ and $a_j$ both conduct the ``REPLAY" operation on the same daily news, the response $O_{j,t}$ from the agent $a_j$ can be treated as a reference stream fed to the agent $a_i$ for interaction.  
Therefore, $\mathcal{V}_{i,t,k}$ is formulated to aggregate the response of agents from group $\mathcal{G}_{t,k}$ having the same operation as $a_i$ with Eq.~\eqref{eq:7},  
\begin{equation}
\mathcal{V}_{i,t,k}
=
\left\{
\mathcal{B}_{j,t}
\;\middle|\;
{a}_j \in \mathcal{G}_{t,k},
\ \operatorname{rank}(S^{rec}_{i,j,t}) \le \tau
\right\},
\label{eq:7}
\end{equation}
where $S_{i,j,t}^{rec}$ denotes the recommendation score between agent $a_j$ and agent $a_i$ at time $t$ within the news-driven group $\mathcal{G}_{t, k}$. 
Moreover, agents are ranked in descending order with the top-$\tau$ selected.
Specifically, $S_{i,j,t}^{rec}$ is calculated based on the content alignment term and the timeliness factor with Eq.~\eqref{eq:8},
\begin{equation}
S_{i,j,t}^{rec} = \underbrace{\Phi(\mathcal{B}_{i,t}, \mathcal{B}_{j,t})}_{\text{content alignment}} \cdot \underbrace{e^{-\beta(t - t_{p})}}_{\text{timeliness factor}} ,
\label{eq:8}
\end{equation}
where the first term is the content alignment term that measures the similarity between the response of $\mathcal{B}_{i,t}$ and $\mathcal{B}_{j,t}$ with function $\Phi(\cdot)$.
The second term is the timeliness factor, which follows an exponential decay with rate $\beta$ between the current time $t$ and the response post time $t_{p}$.
Besides of $\mathcal{R}_{i,t,k}$, the other aspect in Eq.~\eqref{eq:4} is how to update the memory $\boldsymbol{M}_{i,t}$ after interaction. 
Formally, we construct a dynamic memory module to store the agent's opinion history at simulation step $t$, which comprises of the self-reflective memory $\mathcal{SF}(\mathcal{M}_{i,t})$ to record the agent's historical attitude and opinion, and dynamic interaction data $\mathcal{R}_{i,t,k}$. 
The update of individualized memory is formulated as,
{
\setlength{\abovedisplayskip}{8pt}
\setlength{\belowdisplayskip}{8pt}
\begin{equation}
\mathcal{M}_{i,t,k} =\left\{\mathcal{SF}(\mathcal{M}_{i,t}), \mathcal{R}_{i,t,k}  \right\},
\end{equation}%
}
where $\mathcal{SF}(\mathcal{M}_{i,t})$ is derived via the function $\mathcal{SF}(\cdot)$ that dynamically queries and synthesizes the agent's cumulative history to extract historical opinion. More details are provided in the Appendix.

The interaction topology is dynamic. 
As the simulation progresses to the next  $(t+1)$-th step, the introduction of fresh dual-stream news, $\mathcal{D}_{i,t+1}$, drives agents to re-cluster into new news-driven groups based on their interests. 
After several iterative simulations, the ES-MAS enables an efficient simulation of U.S. attitude changes toward China, revealing how dynamic social interactions influence the evolution of public attitudes and collective behaviors, is formalized in Algorithm~\ref{alg:es_mas}.

\begin{table}
    \centering
    \footnotesize
    \caption{Result of macro alignment evaluation. \textbf{Bold} indicates the best results, and \underline{underlining} indicates the second-best results. The $\downarrow$ symbol implies lower is better.}
    \label{tab:main_comparison}
    
    
    \resizebox{0.98\linewidth}{!}{
        \begin{tabular}{l|cccc} 
            \toprule
            & \multicolumn{4}{c}{\textbf{U.S. Attitude Changes Toward China}} \\
            
            Method & $\Delta$Bias $\downarrow$ & $\Delta$Diversity $\downarrow$ & DTW $\downarrow$ & Frechet $\downarrow$ \\ 
            \midrule
            
            BC & 5.5644 & 1.9248 & 22.7863 & 6.7223 \\
            HK & 3.9369 & 1.7856 & 17.3651 & 7.5510 \\
            LR & 2.1805 & 1.6224 & 10.9095 & 3.5467 \\
            SJ & 3.0801 & 1.3135 & 12.8437 & 4.7100 \\
            RA & 1.6639 & 1.2397 & 8.8880 & 3.0467 \\
            \midrule 
            
            FPS & \underline{1.4987} & 1.1945 & \underline{7.2168} & \underline{2.0020} \\
            HiSim & 1.8936 & \underline{1.0445} & 8.9016 & 5.0950 \\
            SOD & 4.1018 & 1.2653 & 16.0846 & 5.1000 \\
            \midrule 
            
            \textbf{ES-MAS} & 
            \textbf{0.5767} & 
            \textbf{0.7423} & 
            \textbf{2.4403} & 
            \textbf{1.6150} \\ 
            \bottomrule
        \end{tabular}
    }
\end{table}

\begin{figure}
    \centering
    \includegraphics[width=1\linewidth]{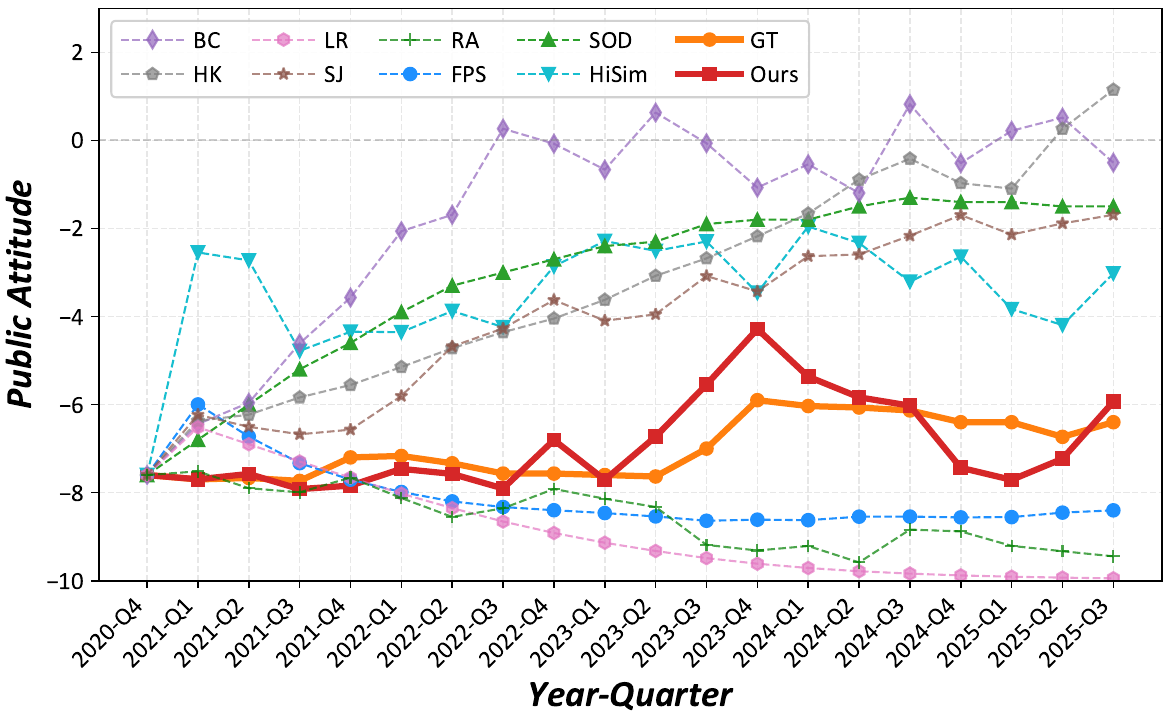} 
    \caption{Comparison of simulated and real-world U.S. public attitude dynamics toward China over time.}
    \label{fig:attitude_comparison}
\end{figure}

\section{Experiment}
\subsection{Experimental Settings}
\textbf{Configurations:} 
We utilize the GPT-4o-mini model for agent reasoning and the text-embedding-3-large model for generating embeddings, with a memory decay rate of $\beta = -0.1$ and a recommendation list size of $\tau = 8$. 
The simulation involves $N = 100$ agents across $T = 20$ quarterly steps, spanning from 2021 to 2025. 
The dataset consists of 258 significant events distributed throughout this timeline. 
During each step, agents process all concurrent significant events and autonomously select $K = 24$ daily news articles from a randomized candidate pool of $\mathcal{N} = 100$.
In total, our simulator processes over 70,000 agent-news interactions.

\textbf{Metrics:} 
To assess the accuracy of ES-MAS in simulating U.S. Attitude Changes Toward China, we evaluate the simulation across two dimensions: 1) \emph{Numerical Alignment}: We use $\Delta\text{Bias}$ and $\Delta\text{Div}$ to quantify the mean deviation and variance stability between simulated and real-world attitudes ~\cite{muller2007information}; 2) \emph{Temporal Dynamics}: To assess how well the simulation captures historical evolutionary trends, we employ Dynamic Time Warping (DTW) ~\cite{eiter1994computing} and Fréchet distance ~\cite{santos2021link};

\textbf{Baselines:} We compare the proposed method with the following baselines: 1) \emph{Agent-Based Models:} BC ~\cite{deffuant2000mixing}, HK ~\cite{hegselmann2015opinion}, RA ~\cite{deffuant2002can}, SJ ~\cite{jager2005uniformity}, and LR ~\cite{lorenz2021individual} which update attitudes via fixed mathematical rules regarding neighbor selection and message acceptance.
2) \emph{LLM-based social network simulators:} FPS ~\cite{liu2024skepticism}, which models belief confirmation and consensus-building processes; SOD ~\cite{chuang2024simulating}, which focuses on simulating confirmation bias and consensus; and HiSim ~\cite{mou2024unveiling}, a hybrid architecture that scales up LLM agents by integrating them with traditional ABM mechanisms.

\begin{table}
    \centering
    \caption{Ablation Analysis of the component in ES-MAS.}
    \label{tab:ablation}
    
    \footnotesize 
    
    \setlength{\tabcolsep}{2.5pt} 
    \renewcommand{\arraystretch}{1.2} 
    
    \begin{tabular}{l|ccc|cccc} 
        \toprule
        & \multicolumn{3}{c|}{\textbf{Components}} & \multicolumn{4}{c}{\textbf{Performance}} \\
        Method & SEIM & PAIM & NDDI & $\Delta$Bias $\downarrow$ & $\Delta$Div. $\downarrow$ & DTW $\downarrow$ & Frech. $\downarrow$ \\ 
        \midrule
        
        \rowcolor{gray!15}
        \textbf{ES-MAS} & \checkmark & \checkmark & \checkmark & \textbf{0.5767} & \textbf{0.7423} & \textbf{2.4403} & \textbf{1.6150} \\ 
        
        \midrule
        w/o DSDIE & $\times$ & $\times$ & $\checkmark$ & 1.4984 & 0.7999 & 6.1764 & 2.0650 \\
        
        Only PAIM & $\times$ & $\checkmark$ & $\times$ & 1.4529 & 1.7247 & 6.7676 & 3.4300 \\

        w/o SEIM & $\times$ & $\checkmark$ & $\checkmark$ & 1.2386 & 1.3714 & 5.2251 & 2.8650 \\

        Only SEIM & $\checkmark$ & $\times$ & $\times$ & 1.1613 & 1.4073 & 4.9091 & 2.6550 \\

        w/o PAIM & $\checkmark$ & $\times$ & $\checkmark$ & 1.0994 & 0.9710 & 4.2318 & 2.5275 \\

        w/o NDDI & $\checkmark$ & $\checkmark$ & $\times$ & 0.9177 & 1.1642 & 3.7478 & 2.5000 \\

        \bottomrule
    \end{tabular}
\end{table}
\subsection{Comparison with State-of-the-Art Methods}
To assess the alignment of U.S. attitudes toward China with reality, we systematically conduct a Macro Alignment Evaluation by comparing the simulated results of ES-MAS with various static baselines. 
As shown in Table \ref{tab:main_comparison}, ES-MAS consistently demonstrates the best performance across all metrics, highlighting its accuracy, robustness, and generability in alignment. 
In contrast, existing simulators that rely on fixed interaction rules and pre-existing static data struggle to accurately model public attitudes toward China. 
This further emphasizes the importance of incorporating real-time news data and the dynamic interactions among agents.

Specifically, regarding Numerical Distribution, ES-MAS achieves a reduction in the mean values of the second-best model by 0.922 $\Delta$Bias and 0.3022 $\Delta$Diversity, thereby substantiating its advantage in marked stability and accurate reflection of public attitude dynamics over long-time simulations. 
This superior performance indicates that the simulated trajectories generated by ES-MAS are tightly synchronized with the real-world timeline. 
Regarding trend shape alignment, ES-MAS exhibits notable improvements in both DTW and Frechet metrics, realizing respective average decreases of 4.7765 and 0.387 relative to the second-best method. 
This affirms its effectiveness in accurately replicating the non-linear fluctuations of historical evolution.
For an intuitive comparison of the differences between the simulated and real-world attitude dynamics, we present the visualization results of the U.S. attitude changes toward China in Fig.~\ref{fig:attitude_comparison}

\subsection{Ablation Study}
In this section, we conduct several ablations to analyze the effect of each component.

\textbf{Dual-Stream Data Integration Engine (DSDIE)}: 
DSDIE integrates both significant events and daily news to provide a continuous, real-time data stream for simulation. 
The setting ``ES-MAS w/o DSDIE'' without considering DSDIE means that the agents are initialized and internal interactions without any reality-grounding updates. 
As shown in Table~\ref{tab:ablation}, it results in the most severe performance degradation, marked by a substantial increase of 0.9217 in $\Delta$Bias and 3.7361 in DTW. 
This shows that continuously injecting dynamic news streams is crucial for accurate simulation, guaranteeing that the simulated public sentiment is consistently influenced by real-world information.

\textbf{Significant Events Inject Module(SEIM)}: 
SEIM directly injects real-time significant events into the simulation environment for establishing a global context accessible to all agents.
By excluding SEIM, agents update attitudes relying solely on the cumulative effect of daily news and internal interactions.
As shown in Table~\ref{tab:ablation}, the setting ``w/o SEIM'' results in a distinct performance decline, \emph{e.g.,} obtaining worse performance of 1.2386, 1.3714, 5.2251, and 2.8650 in terms of $\Delta$Bias, $\Delta$Div., DTW, and Frech., respectively.
This indicates that the injection of significant events plays a crucial role in accurate U.S. Attitude Changes Toward China, as daily news alone fails to capture the critical historical turning points and shifts in public sentiment.

\begin{figure}
    \centering
    \includegraphics[width=1\linewidth]{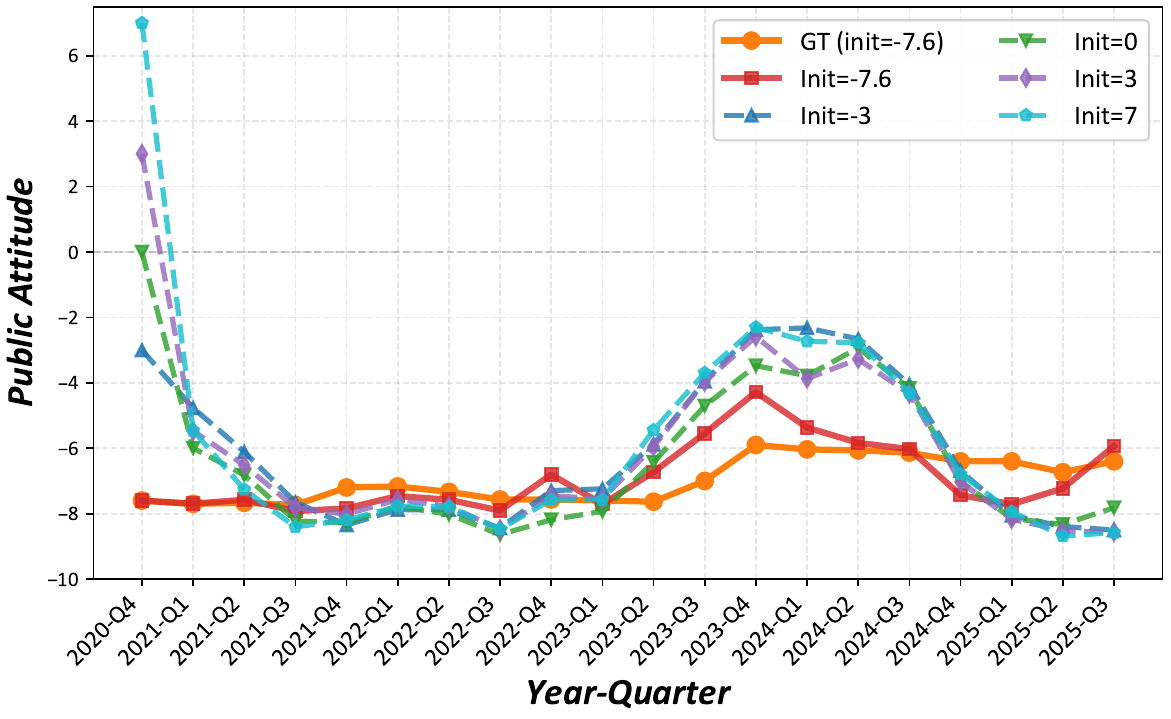} 
    \caption{Sensitivity Analysis on different initial attitudes.}
    \label{fig:sensitivity} 
\end{figure}
\textbf{Personalized Autonomous Ingestion Module (PAIM)}: 
Operating within the context of significant event injection, PAIM constructs a daily news pool that enables agents to autonomously select news matching their preferences.
By discarding PAIM, agents are restricted to reacting solely to significant events without considering their personalized information intake. 
Consequently, they are driven entirely by macro trends, rendering them devoid of independent opinion evolution and prone to artificial consensus.
As shown in Table~\ref{tab:ablation}, the removal of PAIM results in a performance drop, causing an increase of 0.5227 in $\Delta$Bias and 1.7915 in DTW.
This indicates that the autonomous selection from the daily news pool is essential for cognitive diversity in the evolution of China-U.S. relations.

\textbf{News-Driven Dynamic Interaction Module (NDDI)}: 
NDDI aims to collect agents into local groups for facilitating dynamic interactions among agents belonging to the same group.
Once excluding this NDDI from the ES-MAS, agents' interaction becomes trapped in information cocoons because of the lack of information exchange, rendering them unable to adapt to the evolution of the China-U.S. relations.
As shown in Table~\ref{tab:ablation}, the setting ``w/o NDDI" results in a distinct decline, marked by an increase of 0.3410 in $\Delta$Bias and 0.4219 in $\Delta$Diversity.
This indicates that local interactions within dynamic clusters are the driving force behind the emergence of macro-level trends. 

\textbf{Effect of Significant Events}: 
Significant events represent the impact of macro-level dynamic trends. In the Only SEIM setting, the simulation is driven exclusively by significant events, where agents evolve solely based on macro-level shocks, devoid of personalized daily news and dynamic interactions.
As shown in Table~\ref{tab:ablation}, this setting yields a relatively robust performance, achieving a $\Delta$Bias of 1.1613. 
This result outperforms the w/o SEIM setting (1.2386) in terms of bias alignment, suggesting that significant events are the primary determinant of the overall attitude baseline. 
This demonstrates that {SEIM} successfully captures the fundamental turning points and the structural trend of China-U.S. relations, ensuring the macro-direction remains correct.

\textbf{Effect of Daily News}: 
In the Only PAIM setting, agents are exposed solely to daily news, meaning their opinion evolution is driven exclusively by the autonomous selection of information, lacking both macro-level guidance and dynamic interactions among agents. 
As shown in Table~\ref{tab:ablation}, this setting suffers from significant misalignment, yielding a $\Delta$Bias of 1.4529 and a DTW of 6.7676.
Notably, this performance is inferior to the Only SEIM setting ($\Delta$Bias 1.1613), this indicate that agents tend to get lost in diverse daily information.
Therefore, the daily news provides the necessary details for preventing the opinion trajectory from reality.
\begin{figure*}
    \centering
    \includegraphics[width=1\linewidth]{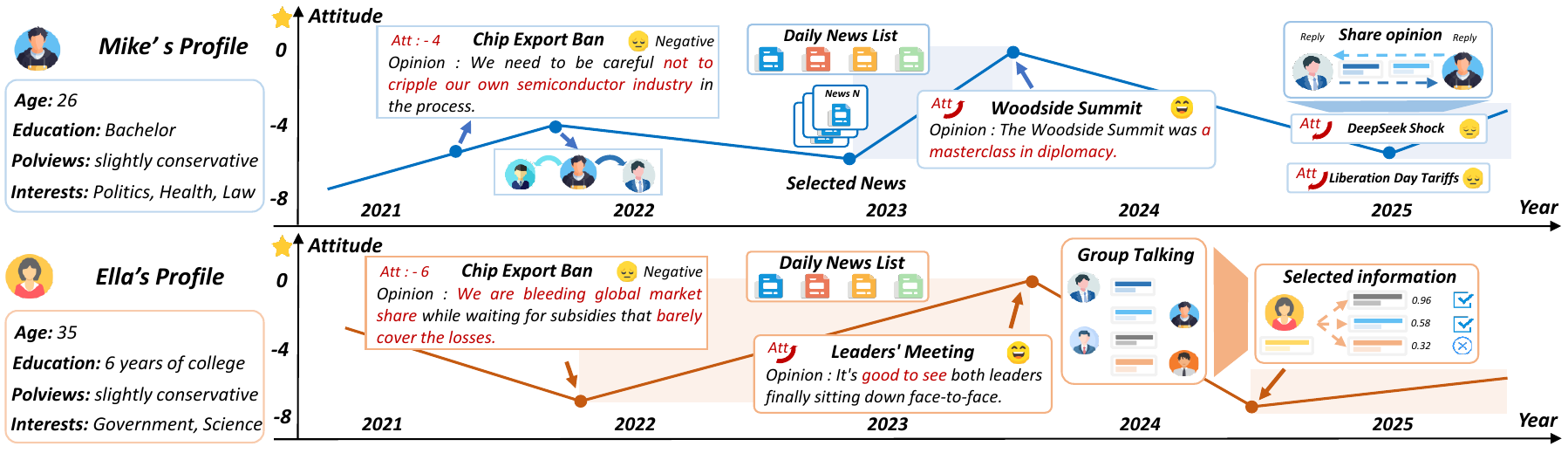}
    \caption{The attitude trajectories of two representative agents (Mike and Ella) toward China from 2021 to 2025, illustrates how individual opinions dynamically adapt to macro-level geopolitical shocks (e.g., the Chip Export Ban and Woodside Summit) and micro-level influences (e.g., personalized news selection and dynamic peer interactions). }
    \label{fig:case}
\end{figure*}

\textbf{Effect of Initial Attitudes}: 
The initial attitude is the initial opinion of each agent about the U.S. Attitudes Toward China, while a different attitude would cause a different simulation trend. 
Therefore, a good simulator should be insensitive to different initializations.
To evaluate the robustness of ES-MAS, we conducted a sensitivity analysis with a variety of initial attitudes. 
Formally, we initialized the agents with five distinct mean attitude values, \emph{i.e.,} $-7.6, -3, 0, 3, 7$. 

As shown in Figure~\ref{fig:sensitivity}, ES-MAS exhibits a consistent simulation trend for the distinct initial attitudes, thereby successfully reproducing the macro-level evolutionary trend.




\textbf{Effect of Micro-level}:
At the individual cognitive level, Figure~\ref{fig:case} tracks the attitude trajectories of two representative agents, Mike and Ella, to illustrate how our framework's mechanisms drive opinion shifts. 
When encountering macro-level shocks and micro-level influences, the agents generate distinct attitude fluctuations supported by personalized rationales. 
This visualization clearly illustrates how individual opinions realistically adapt to a variety of complex exogenous events and endogenous social interactions over time. 
Consequently, this micro-level analysis validates the internal mechanisms within ES-MAS, demonstrating its effectiveness in driving macro-level trends through bottom-up individual behaviors.

\subsection{Scalability Analysis}
To evaluate the scalability of the proposed ES-MAS, we conduct further evaluation based on varying the agent numbers $N \in \{25, \allowbreak 50, \allowbreak 100, \allowbreak 200, \allowbreak 500\}$.
Moreover, we compare ES-MAS against the second-best FPS across four alignment metrics, and summarize the related results in Table \ref{tab:scalability}. 
It can be seen that increasing the number of agents can improve the accuracy of the simulation. 
However, when the number of agents exceeds 100, the accuracy of the simulation decreases.
The reason is that an excessive number of agents introduces complex interaction dynamics, which facilitates the emergence of more polarized opinions and impedes the agents' to accurately judge the current real-time situation.

\begin{table}
    \centering
    \caption{Scalability analysis regarding FPS (baseline) and ES-MAS.}
    \label{tab:scalability}
    
    \footnotesize
    
    \setlength{\tabcolsep}{4pt}
    \renewcommand{\arraystretch}{1.2}
    
    \begin{tabular}{lc|cccc} 
        \toprule
        \multicolumn{2}{c|}{} & \multicolumn{4}{c}{\textbf{Scalability Analysis (N agents)}} \\
        Method & Size ($N$) & $\Delta$Bias $\downarrow$ & $\Delta$Diversity $\downarrow$ & DTW $\downarrow$ & Frechet $\downarrow$ \\ 
        \midrule
        
        \multirow{5}{*}{\textbf{ES-MAS}} 
        & 25  & 0.8388 & 1.0134 & 3.5158 & 1.6900 \\
        & 50  & 0.9783 & 1.1771 & 4.2899 & 2.3300 \\
        & 100 & \textbf{0.5767} & \textbf{0.7423} & \textbf{2.4403} & \textbf{1.6150} \\
        & 200 & 1.5027 & 1.7857 & 7.1429 & 3.0050 \\
        & 500 & 1.3275 & 1.6052 & 6.2304 & 3.5056 \\
        \midrule 

        \multirow{5}{*}{FPS} 
        & 25  & 1.3650 & 1.4309 & 6.7985 & 2.1800 \\
        & 50  & 1.2461 & 1.3043 & 6.3538 & 1.9900 \\
        & 100 & 1.4987 & 1.1945 & 7.2168 & 2.0020 \\
        & 200 & 6.0051 & 1.8115 & 25.3306 & 9.1530 \\
        & 500 & 6.3244 & 1.6678 & 28.5097 & 8.8335 \\     
        \bottomrule
    \end{tabular}

\end{table}

Furthermore, compared to the existing FPS simulator, the proposed ES-MAS demonstrates strong performance across various scales, with minimal differences in performance regardless of the number of agents. 
For instance, in terms of numerical stability, ES-MAS consistently maintains low error rates across all four metrics, even as the number of agents, $N$, increases to 500. 
When $N$ reaches 200, FPS experiences a significant spike in $\Delta$Bias, measuring 6.0051, whereas ES-MAS retains a much lower $\Delta$ Bias of 1.5027. At $N = 500$, FPS fails to converge, showing a $\Delta$ Bias greater than 6.3, while ES-MAS remains stable with a $\Delta$ Bias of 1.3275. 
This demonstrates that our simulator effectively reduces the collective polarization typically observed in scaling Multi-Agent Systems through dynamic interactions among agents and real-time news data.

Regarding trend shape alignment, our simulator maintains the temporal accuracy of the simulation, while FPS experiences a severe decline. 
At $N = 500$, the Dynamic Time Warping (DTW) for FPS skyrockets to 28.5097, indicating a complete disconnection from the historical timeline. In contrast, ES-MAS achieves a low DTW of 6.2304. 
This suggests that ES-MAS ensures that macro-level evolutionary trends are kept closely aligned with reality, even as the agent population increases.



\section{Conclusion}
In this work, we propose ES-MAS, an Event-Steered Multi-Agent Simulator designed to understanding the evolution of U.S. public attitudes toward China. 
By integrating dual-stream data and news-driven dynamic interactions among agents, ES-MAS ensuring a degree of alignment between simulated societal trends and historical reality.
Unlike traditional approaches that rely on static rules and data, our framework effectively captures the continuous, bottom-up emergence of public opinion. 
Extensive experiments demonstrate that ES-MAS offers a approach for analyzing complex geopolitical dynamics, providing insights into the mechanisms driving international relations.

\section*{Ethical Statement}
The development and application of the Event-Steered Multi-Agent Simulator (ES-MAS) require careful ethical consideration, particularly given its focus on sensitive geopolitical topics. 
To prevent misuse for manipulating public opinion, transparency and strict supervision are essential, necessitating standardized usage scopes, clear permissions, and defined ethical boundaries.
Regarding data usage, the CURE dataset is constructed entirely from publicly available news sources and historical event records (e.g., The Guardian API, policy briefings). 
No personally identifiable information or private user data was involved in the simulation process. We are committed to ensuring data reliability while strictly respecting privacy boundaries.
To mitigate this, robust verification and control mechanisms will be implemented to prevent the spread of harmful information. By adhering to stringent standards and safeguards, ES-MAS will be used solely for legitimate purposes, avoiding any form of information manipulation or illegal activities.

\bibliographystyle{IEEEtran}
\bibliography{reference}

\end{document}